\documentclass[aps,prl,reprint]{revtex4-1}
\usepackage{graphicx}
\usepackage{times}
\usepackage{bm}
\usepackage{amsmath,amssymb}
\usepackage{feynmf}
\usepackage{multirow}

\begin{document}
\title{Some features of the direct and inverse double Compton effect as applied to astrophysics}
\author{V. Dubrovich$^{2}$, T. Zalialiutdinov$^{1,2}$}
\affiliation{
$^1$ Department of Physics, St. Petersburg State University, Petrodvorets, Oulianovskaya 1, 198504, St. Petersburg, Russia\\
$^2$ Special Astrophysical Observatory, St. Petersburg Branch, Russian Academy of Sciences, 196140, St. Petersburg, Russia
}
\email[E-mail:]{t.zalialiutdinov@spbu.ru}

\begin{abstract}
In the present paper, we consider the process of inverse double Compton (IDC) scattering in the context of astrophysical applications. It is assumed that the two hard X-ray photons emitted from an astrophysical source are scattered on a free electron and converted into a single soft photon of optical range. Using QED S-matrix formalism for the derivation of a cross-section of direct double Compton (DDC) and assuming detailed balance conditions we give an analytical expression for the cross-section of the IDC process. It is shown that at fixed energies of incident photons the inverse cross-section has no infra-red divergences and its behavior is completely defined by the spectral characteristics of the photon source itself, in particular, by the finite interaction time of radiation with an electron. Thus, even for the direct process, the problem of resolving infrared divergence actually refers to a real physical source of radiation in which photons are never actually plane waves.  As a result the physical frequency profile of the scattered radiation for direct as well as inverse double Compton processes is a function of both the intensity and line shape of the incident photon field.
\end{abstract}

\maketitle

The object of this paper is to give an expression for the inverse double Compton effect (IDC), i.e. the third-order process in which two hard photons colliding with a free electron give rise to one scattered soft photon. The interest of such calculations lies in the fact that, though the cross-section for this process is small in comparison with ordinary Compton scattering, nevertheless, the high-intensity X-ray photons in the vicinity of astrophysical sources allow a possible study of these phenomena.

The ordinary direct double Compton scattering has been repeatedly considered by many authors \cite{HeitlerNordheim, Elizier, Cavanagh, LEVINE1971197}. The exact relativistic expressions for the differential cross-section of the DC process accounting angular correlations were obtained in \cite{mandl1952} within the QED formalism of S-matrix. Various proposals for laboratory studies of the double Compton effect have also been actively discussed in recent decades \cite{jent1,jent2, Sherwin}. In addition, astrophysical applications are also of particular interest when describing the spectral features of various astrophysical sources and comparing them with ground-based or satellite observations \cite{Sunyaev}. 

The inverse double Compton scattering is also of particular importance. Under astrophysical conditions, IDC can be important for objects with a very high density of photons (for example X-ray sources) or the in the early universe at a high intensity of the CMB. In these cases, the intensity of the generated high-frequency Raman radiation can be noticeable. For example, in the presence of a powerful line in the spectrum of the X-ray shell, in the observed spectrum there will be an admixture of a line with a doubled energy, which cannot be identified with any chemical element. The importance of taking this effect into account is determined by the accuracy with which it is necessary to describe the physical parameters of the source and to avoid false identifications and estimates. Processes of this kind in various astronomical applications have been considered earlier \cite{Lightman}. In particular, in  \cite{Lightman, chluba2} the inverse processes included in the equations of balance and radiation transfer were expressed through direct ones, following the principle of detailed balance. However, as will be shown below, the forward process has divergence in the infrared region that are absent in the reverse process if we do direct calculations.

When considering the direct double Compton effect, there is some subtle effect of a fundamental nature - a feature of the formation of the infrared (low-frequency) end of the spectrum of the resulting photons - the so-called infrared catastrophe \cite{chluba2}. In the literature, there are a large number of versions of the solution to this problem \cite{mandl1952,19}. To consistently cure the issue, the DC process has to be treated together with the next to leading order (NLO) radiative corrections to the Compton process \cite{feynman, chluba2}. In this context, one is usually interested in the total Compton scattering cross-section at order $\alpha^3$. NLO corrections display a logarithmic divergence which can be shown to cancel with the corresponding DC one. Employing a regularization — introducing a photon mass in the standard approach \cite{feynman}  - in the calculation of both cross-sections and summing the two, the dependence on the regularization parameters drops out, leaving a finite radiative correction at order  $\alpha^3$. The drawback of this approach is that the radiative correction cross-section now depends on the energy resolution of the experiment, $\omega_{\mathrm{res}}$. The argument to justify summing the two processes is that below some $\omega_{\mathrm{res}}\ll 1$ the experiment is unable to distinguish contributions from virtual photon emission, relevant to the computation of the NLO correction, and from real photon emission by DC. Adding all contributions to the total CS scattering cross-section it was found that the correction can exceed the naive $\alpha/\pi$ level at sufficiently high energies \cite{Mork}. It is should be noted that direct and inverse process were recently studied in the laboratory conditions \cite{India}. In \cite{India} it was shown that measured DDC and IDC cross-sections are of comparable magnitudes and agrees with theoretical predictions. From this, it can be concluded that it is important to take into account both processes in problems related to radiation transfer.

However, all of the regularization procedures are related to the consideration of the final state of the system \cite{feynman}. In this work, we propose a different physical approach for solving this problem based on the analysis of the input photon - taking into account its finiteness in time. Indeed, the standard consideration assumes that the incident photon is taken in the form of a plane wave infinite in time and space \cite{Akhiezer, bjorken1964relativistic, Landau, Berest}. Only in this case there is possibility to form an arbitrarily low-frequency photon at the end of the process. However, in reality, there are no such waves in nature. There are only real photons, which are formed by some radiation mechanism during a finite time interval \cite{dawson1970}. As a result, the scattering of such a photon by an electron lead to a finite spectrum that is completely finite in the entire frequency range. From the point of view of observational astrophysics, this provides a alternative way to determine the parameters of an incident photon from the spectrum of the low-frequency wing. Thus, we obtain a new independent channel of information about the mechanism of radiation formation in the source.

Below we will show that in real physical conditions the problem of eliminating the infrared divergence in the cross-section of the direct process refers to the finite interaction time of radiation with an electron, which can be introduced at least phenomenologically within the QED formalism. For these purposes, we first consider the cross-section of the reverse process in the context of its astrophysical application. We assume that two photons emitted in the vicinity of some object are scattered on a free electron and convert into a single photon $ \gamma(k_1)+\gamma(k_2) +e^{-}\rightarrow  \gamma(k) + e^{-} $.  
It is assumed that such a process, along with a direct one, can leave significant distortions in the spectrum of some astrophysical source in the vicinity of which events of collisions of photons with an electron gas occur. The characteristic behavior of the cross-section of inverse double Compton scattering is derived within the framework of the rigorous QED approach.

To begin with, we recall a brief derivation of the scattering cross section of the direct process $ \gamma(k)+e^{-}\rightarrow \gamma(k_1)+\gamma(k_2) + e^{-} $ \cite{Melrose1972}.
Following \cite{mandltextbook} the direct double Compton (DDC) is described by the third order S-matrix element (here and below we use relativistic units $ \hbar=c=1 $, $ \alpha=e^2/4\pi $, the electron mass $ m_{e} $ is written explicitly)
\begin{eqnarray}
\label{mandl1}
S^{(3)}_{fi}=(2\pi)^4\delta^{(4)}(p_{f}+k_1+k_2-p_{i}-k)
\\\nonumber
\times
\frac{m_{e}e^3}{\sqrt{8 V^5 E_{i}E_{f}\omega  \omega _{1}\omega _{2}}}{\cal M}(k,k_1,k_2)
,
\end{eqnarray}
where $ p_{i}=(E_{i},\textbf{p}_{i}) $, $ p_{f}=(E_{f},\textbf{p}_{f}) $  are the four-vectors of initial and final electron momenta, respectively; $ k=(\omega,\textbf{k}) $ is the four vector of incident photon with frequency $\omega$ and wave-vector $ \textbf{k} $, $ k_{1}=(\omega_{1},\textbf{k}_{1}) $ and $ k_{2}=(\omega_{2},\textbf{k}_{2}) $  are the four-vectors of two outgoing photons, and  $ \cal{M} $ is the Feynman amplitude of the process 
\begin{widetext}
\begin{eqnarray}
\label{mandl2}
{\cal M}(k,k_1,k_2)=
\overline{u}(p_{f})
\left\lbrace
\hat{\varepsilon}(k)\frac{\hat{p}_{f}-\hat{k}+m_{e}}{(k-p_{f})^2-m_{e}^2}\hat{\varepsilon}^*(k_{1})
\frac{-\hat{k}+\hat{k}_1+\hat{p}_{f}+m_{e}}{(k-k_1-p_{f})^2-m_{e}^2}
\hat{\varepsilon}^*(k_{2})
+(1\leftrightarrow 2)
\right.
\\\nonumber
\left.
+
\hat{\varepsilon}^*(k_1)\frac{\hat{k}_1+\hat{p}_f+m_{e}}{(-k_1-p_{f})^2-m_{e}^2}\hat{\varepsilon}(k)
\frac{-\hat{k}+\hat{k}_1+\hat{p}_{f}+m_{e}}{(k-k_1-p_{f})^2-m_{e}^2}
\hat{\varepsilon}^*(k_{2})
+(1\leftrightarrow 2)
\right.
\\\nonumber
\left.
+
\hat{\varepsilon}^*(k_1)\frac{\hat{k}_1+\hat{p}_f+m_{e}}{(-k_1-p_{f})^2-m_{e}^2}\hat{\varepsilon}^*(k_2)
\frac{\hat{k}_1+\hat{k}_2+\hat{p}_{f}+m_{e}}{(-k_1-k_2-p_{f})^2-m_{e}^2}
\hat{\varepsilon}(k)
+(1\leftrightarrow 2)
\right\rbrace
u(p_{i})
.
\end{eqnarray}
\end{widetext}
Here $u(p)$ is the Dirac spinor for free electron with its Dirac adjoint defined as $ \overline{u}(p)=u^{\dagger}(p)\gamma_{0} $, $ \gamma^{\mu} $ ($ \mu=0,\,1,\,2,\,3 $) are the Dirac matrices and $\varepsilon(k)$ is the polarization four-vector of a photon $ k $.
The contraction of $ \gamma^{\mu} $ matrices with four-vector $ a $ is given by $ \hat{a}=\gamma^{\mu}a_{\mu}$.

The transition rate per unit time to one define state can be found according to the definition 
\begin{eqnarray}
\label{mandl3}
w=\frac{|S^{(3)}_{fi}|^2}{T}=V(2\pi)^4\delta^{(4)}(p_{f}+k_1+k_2-p_{i}-k)
\\\nonumber
\times
\frac{m_{e}^2e^6}{8 V^5 E_{i}E_{f}\omega  \omega _{1}\omega _{2}}|{\cal M}(k,k_1,k_2)|^2
,
\end{eqnarray}
where $ T\rightarrow \infty $ is the observation time and $ V $ is the phase volume. Since we are interested to a transition rate $ dw $ to a group of final states with momenta in the intervals $ (\textbf{p}_f,\textbf{p}_f + d\textbf{p}_f) $, $ (\textbf{k}_1,\textbf{k}_1 + d\textbf{k}_1) $ and $ (\textbf{k}_2,\textbf{k}_2 + d\textbf{k}_2) $ we must multiply Eq. (\ref{mandl3}) by the number of these states which is 
\begin{eqnarray}
\label{mandl4}
\frac{V^3d^3\textbf{p}_{f}d^3\textbf{k}_1 d^3\textbf{k}_2}{(2\pi)^9}
.
\end{eqnarray}
With our choice of normalization for the states, the volume $ V $ contains one scattering center and the incident photon flux is  $ F=c/V $ ($c$ is the speed of light) \cite{rapoport}. Then the corresponding differential DDC cross-section can be found as follows
\begin{eqnarray}
\label{mandl5}
d\sigma_{\mathrm{ddc}}=\frac{dw}{F}
=\delta^{(4)}(p_{f}+k_1+k_2-p_{i}-k)
\\\nonumber
\times
\frac{m_{e}^2e^6}{8E_{i}E_{f}\omega  \omega _{1}\omega _{2}}|{\cal M}(k,k_1,k_2)|^2 \frac{d^3\textbf{p}_{f}d^3\textbf{k}_1 d^3\textbf{k}_2}{(2\pi)^5}
.
\end{eqnarray}

Integration over final electron momenta in Eq. (\ref{mandl5}) can be written in a four dimensional form with the use of equality \cite{bjorken1964relativistic}
\begin{eqnarray}
\label{mandl6}
\frac{d^3\textbf{p}_f}{2E_f}=\int d^4p_f \delta(p_f^2-m_e^2)\theta(p_{f}^0)
,
\end{eqnarray}
where $ m_{e}^2=E_{f}^2-\textbf{p}_{f} $, $\theta(p_{f}^0)$ is the Heaviside step function. Finally, performing  integration over $ d^4p_{f} $ with the use delta function properties we find \cite{mandl1952,jent2}
\begin{eqnarray}
\label{mandl7}
d\sigma_{\mathrm{ddc}}
=\delta((p_{i}+k-k_1-k_2)^2-m_e^2)
\\\nonumber
\times
\frac{m_{e}^2e^6}{4E_{i}\omega \omega_{1}\omega_{2}}|{\cal M}(k,k_1,k_2)|^2 \frac{d^3\textbf{k}_1 d^3\textbf{k}_2}{(2\pi)^5}
.
\end{eqnarray}

We now specialize to a reference frame where the electron initially at rest, i.e. $ \textbf{p}_{i}=0 $ and $E_{i}=m_{e}$. Then the argument of delta-function in Eq. (\ref{mandl7}) becomes
\begin{eqnarray}
\label{mandl8}
(p_{i}+k-k_1-k_2)^2-m_e^2
\\\nonumber
=2 m_{e} (\omega-\omega_1-\omega_2)
-2 \omega \omega_1(1-\cos \chi_1)
\\\nonumber
-2 \omega \omega_2(1-\cos \chi_2)+2 \omega_1 \omega_2(1-\cos \chi_{12})=0
,
\end{eqnarray}
where $ \chi_1 $ is the angle between vectors $\textbf{k} $ and $\textbf{k}_1 $,  $ \chi_2 $ is the angle between vectors $\textbf{k} $ and $\textbf{k}_2 $, and $ \chi_{12} $ is the angle between vectors $\textbf{k}_1 $ and $\textbf{k}_2 $. Here we assumed that $ p_{i}^2=m_{e}^2 $, and $ k^2=k_{1}^2=k_{2}^2=0 $. Then the energy conservation law has the form
\begin{eqnarray}
\label{mandl9}
\omega_2=\frac{m_{e} (\omega- \omega_{1})-\omega \omega_{1}(1-\cos \chi_1)}{m_{e}+\omega(1- \cos \chi_2)-\omega_{1}(1-\cos
   \chi_{12})}
.
\end{eqnarray}

Taking into account that in spherical coordinates $ d^3\textbf{k}_{1(2)}=\omega_{1(2)}^2d\omega_{1(2)}d\Omega_{1(2)} $ (where $ d\Omega_{1(2)}=\sin\theta_{1(2)} d\theta_{1(2)} d\phi_{1(2)} $ and $ \theta_{1(2)} $, $ \phi_{1(2)} $ are the spherical angles of vector $\textbf{k}_{1(2)}$), performing integration over $ d\omega_{2} $ in Eq. (\ref{mandl7}) 
with the use of equality
\begin{eqnarray}
\label{mandl10}
\int dx \delta(f(x))=\left| \frac{df(x)}{dx}\right|^{-1}
,
\end{eqnarray}
and summing over the photon polarizations and electron spin in initial and final states  we find 
\begin{widetext}
\begin{eqnarray}
\label{mandl11}
d\sigma_{\mathrm{ddc}}
=
\frac{m_{e}e^6}{2^8\pi^5 }\left(\frac{\omega _{1}}{\omega}\right)\frac{m_{e} (\omega- \omega_{1})-\omega \omega_{1}(1-\cos \chi_1)}{(m_{e}+\omega(1- \cos \chi_2)-\omega_{1}(1-\cos
   \chi_{12}))^2}  
  X(\omega, \omega_1) d\omega_1 d\Omega_1 d\Omega_2
,
\end{eqnarray}
\end{widetext}
where 
\begin{eqnarray}
\label{mandl12}
X(\omega, \omega_1) =\frac{1}{4}\mathrm{Tr}\sum\limits_{\mathrm{polarizations}}|{\cal M}(\omega, \omega_1)|^2 
.
\end{eqnarray}
Summation in Eq. (\ref{mandl11}) and algebra with Dirac matrices can be performed in a fully analytical way with the use of {\it FeynCalc}. software \cite{FeynCalc1, FeynCalc2}. 

To describe the stimulated DDC cross-section, magnitude Eq. (\ref{mandl2}) should be also multiplied by the factor $ \sqrt{N_{\textbf{k}\textbf{e}}}\sqrt{N_{\textbf{k}_1\textbf{e}_1}+1}\sqrt{N_{\textbf{k}_2\textbf{e}_2}+1}  $, where $N_{\textbf{k}\textbf{e}}$ is the number of photons with a wave-vector $\textbf{k}$ and polarization $ \textbf{e} $. In this case the corresponding flux of incident photons is $ F=cN_{\textbf{k}\textbf{e}}/V $ \cite{rapoport}. Then the number of incident photons vanishes in cross-section. If the intensity of the $ J_{\textbf{k}\textbf{e}} $ radiation is known then $N_{\textbf{k}\textbf{e}}$ can be obtained with the use of following relation
\begin{eqnarray}
\label{mandl13}
N_{\textbf{k}\textbf{e}}=\frac{8\pi^3 c^2}{\hbar \omega^3}J_{\textbf{k}\textbf{e}}
.
\end{eqnarray}

It is well-known that DDC cross-section Eq. (\ref{mandl11}) has infrared divergence of the type $ \omega_{1}^{-1} $ \cite{mandl1952}. In \cite{feynman} it was shown that accounting of QED radiative corrections lead to the natural cut-off of the lower limit of frequency $\omega_1$ which can be attributed to the minimal resolution of the detector in the experiment. Below we show that for the inverse process the divergence is moved from the problem of resolution detector of photons in final states to the source of initial photons.

Using the definition for transition rate Eq. (\ref{mandl2}) we can write equation for the cross-section of inverse double Compton scattering  (IDC) as follows
\begin{eqnarray}
\label{mandl14}
d\sigma_{\mathrm{idc}}=\frac{dw}{F}
=\delta^{(4)}(p_{f}+k-p_{i}-k_1-k_2)
\\\nonumber
\times
\frac{m_{e}^2e^6}{8E_{i}E_{f}\omega  \omega _{1}\omega _{2}}|{\cal M}(-k,-k_1,-k_2)|^2 \frac{d^3\textbf{p}_{f}d^3\textbf{k} }{(2\pi)^2}
.
\end{eqnarray}

In the reference frame where the electron initially at rest ($ \textbf{p}_{i}=0 $) the frequency of outgoing photon $ \omega $ takes the form
\begin{eqnarray}
\label{mandl15}
\omega=\frac{\omega_1\omega_2(1-\cos \chi_{12})+m_e(\omega_1 + \omega_2)}{m_e+\omega_1(1-\cos\chi_1)+\omega_2(1-\cos\chi_2)}
.
\end{eqnarray}
The scalar products of the four vectors included in Eq. (\ref{mandl15}) are given by the following equalities:
$p_{i}k=m_{e}\omega$,
$p_{i}k_{1}=m_{e}\omega_{1}$,
$p_{i}k_{2}=m_{e}\omega_{2}$,
$kk_{1}=\omega\omega_1(1-\cos\chi_1)$,
$kk_{2}=\omega\omega_2(1-\cos\chi_2)$,
$k_{1}k_{2}=\omega_1\omega_2(1-\cos\chi_{12})$,
$p_{f}k=m_{e}\omega+\omega\omega_1(1-\cos\chi_1)+\omega\omega_2(1-\cos\chi_2)$,
$p_{f}k_{1}=m_{e}\omega-\omega\omega_1(1-\cos\chi_1)+\omega_1\omega_2(1-\cos\chi_{12})$,
$p_{f}k_{2}=m_{e}\omega-\omega\omega_2(1-\cos\chi_2)+\omega_1\omega_2(1-\cos\chi_{12})$ and $ p_{f}p_{i}=m_{e}(\omega_1+\omega_2-\omega)+ m_{e}^2$

Then, the after integration over $ d^3\textbf{p}_f $, $ d\omega $ and summation over the photon polarizations and electron spin in initial and final states we find
\begin{widetext}
\begin{eqnarray}
\label{mandl16}
d\sigma_{\mathrm{idc}}
=
\frac{m_{e}e^6}{2^8\pi^2 }\left(\frac{1}{\omega_1\omega_2}\right)
\frac{\omega_1\omega_2(1-\cos \chi_{12})+m_e(\omega_1 + \omega_2)}{(m_e+\omega_1(1-\cos\chi_1)+\omega_2(1-\cos\chi_2))^2}
  \tilde{X}(\omega_1, \omega_2) d\Omega
,
\end{eqnarray}
\end{widetext}
where 
\begin{eqnarray}
\label{mandl17}
\tilde{X}(\omega_1, \omega_2) =\frac{1}{4}\mathrm{Tr}\sum\limits_{\mathrm{polarizations}}|{\cal M}(-k_1, -k_2)|^2 
.
\end{eqnarray}
The magnitude $ {\cal M} $ for the inverse process in Eq. (\ref{mandl17}) differs from the direct one only in the use of a different expression for the energy conservation law (see Eqs. (\ref{mandl9}) and (\ref{mandl15}) ) and by the replacement $ k \leftrightarrow -k $ , $ k_1 \leftrightarrow -k_1 $ and $ k_2 \leftrightarrow -k_2 $.

Following \cite{mandl1952} one can consider the particular case when particles are moving nearly in forward direction. Then introducing notations $ x=1-\cos\chi_1 $, $ y=1-\cos\chi_2 $  and $ z=1-\cos\chi_{12}$ we find
\begin{eqnarray}
\label{mandl18}
d\sigma_{\mathrm{idc}}=2\pi\alpha r_{0}^2
\left(
\frac{\omega_1^2}{\omega_2^2}+\frac{\omega_2^2}{\omega_1^2}+\frac{2 \omega_1}{\omega_2}+\frac{2 \omega_2}{\omega_1}+3\right)
\\\nonumber
\times
\left(\frac{x}{\omega_2}+\frac{y}{\omega_1}-\frac{z}{\omega_1+\omega_2}\right)
d\Omega
,
\end{eqnarray}
where $ \chi_1 $, $ \chi_2 $ and $ \chi_{12} $ are assumed to be small, $ r_{0}=\frac{e^2}{4\pi m_{e}} =2.8\times 10^{-15}\;\mathrm{m} $  and $ \alpha=\frac{e^2}{4\pi} $ is the fine structure constant. As well as DDC cross-section Eq. (\ref{mandl18}) vanishes for scattering in forward direction , i.e. when $ \chi_1=\chi_2=0 $ (i.e. $ x=y=z=0 $ in eq. (\ref{mandl18})). From Eq. (\ref{mandl18} its is seen that IDC scattering cross-section (as well as DDc) is the order $ \alpha $ smaller than single compton scattering.

\begin{figure}[hbtp]
\caption{The IDC distribution function $ F(\omega_{1},\omega_{2}) $, see Eqs. (\ref{mandl18full}), (\ref{mandl18fullFF})}
\centering
\includegraphics[scale=0.7]{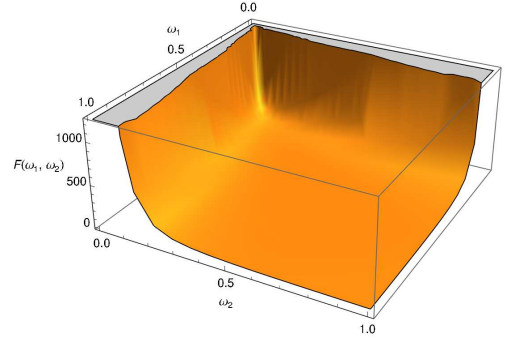}
\label{fig1}
\end{figure}

The dependence on angles $ \chi_{1} $, $ \chi_{1} $, $ \chi_{12} $  in Eq. (\ref{mandl18}) can be expressed in terms of spherical angles ($ \theta $, $ \phi $), ($ \theta_1 $, $ \phi_1 $) and ($ \theta_2 $, $ \phi_2 $) with the use of relations
\begin{eqnarray}
x=1-\cos\phi\cos\phi_1-\cos(\theta-\theta_1)\sin\phi\sin\phi_1
\end{eqnarray}
\begin{eqnarray}
y=1-\cos\phi\cos\phi_2-\cos(\theta-\theta_2)\sin\phi\sin\phi_2 
\end{eqnarray}
\begin{eqnarray}
z=1-\cos\phi_1\cos\phi_2-\cos(\theta_1-\theta_2)\sin\phi_1\sin\phi_2
\end{eqnarray}
For particular case, when two incident photons $ \textbf{k}_1 $ and $\textbf{k}_2$ are propagating parallel to each other, it is convenient to choose $ z $-axis of spherical coordinate system along this direction. Then setting $\theta_1 =\theta_2=0 $,  $\phi_1 =\phi_2=0 $ and performing integration over the $d\Omega = \sin\theta d\theta d\phi$ in Eqs. (\ref{mandl18}) we find
\begin{eqnarray}
\label{mandl18full}
\sigma_{\mathrm{idc}}=8\pi^3\alpha r_{0}^2
F(\omega_{1},\omega_{2}),
\end{eqnarray}
\begin{eqnarray}
\label{mandl18fullFF}
F(\omega_{1},\omega_{2})=\left(
\frac{\omega_1^2}{\omega_2^2}+\frac{\omega_2^2}{\omega_1^2}+\frac{2 \omega_1}{\omega_2}+\frac{2 \omega_2}{\omega_1}+3\right)
\\\nonumber
\times
\left(\frac{1}{\omega_2}+\frac{1}{\omega_1}\right)
\end{eqnarray}
The function $ F(\omega_{1},\omega_{2}) $ is drawn in Fig. \ref{fig1}. It is should be noted that $ \omega_{1} $,  $ \omega_{2} $ are the input parameters of cross-section and they nevet turns to zero for the real source of photons.

For incident photon with equal energies, $ \omega_{1}=\omega_{2} $, Eq. (\ref{mandl18}) turns to
\begin{eqnarray}
\label{mandl19full}
\sigma_{\mathrm{idc}}=\frac{144 \pi^3  \alpha r_0^2}{\omega_1}.
\end{eqnarray}   

In the case IDC the accounting of stimulated scattering can be taken into account by multiplying the magnitude of the process by the factor $ \sqrt{N_{\textbf{k}_1\textbf{e}_1}}\sqrt{N_{\textbf{k}_2\textbf{e}_2}}\sqrt{N_{\textbf{k}\textbf{e}}+1} $ and dividing by the photon flux $ F=c N_{\textbf{k}_2\textbf{e}_2}N_{\textbf{k}_1\textbf{e}_1}/V $ \cite{rapoport}. 
Assuming that for direct process there is a system of electron + photon field with $ N_{\textbf{k}\textbf{e}} $ photons the following relation between
the transition rates of direct and inverse processes can be found \cite{rapoport, LabKlim}
\begin{eqnarray}
\label{tudasuda}
\frac{dw_{\mathrm{ddc}}}{dw_{\mathrm{idc}}}=\frac{N_{\textbf{k}_1\textbf{e}_1}N_{\textbf{k}_2\textbf{e}_2}(N_{\textbf{k},\textbf{e}}+1)}{N_{\textbf{k},\textbf{e}}(N_{\textbf{k}_1\textbf{e}_1}+1)(N_{\textbf{k}_2\textbf{e}_2}+1)}
.
\end{eqnarray}
The equation above is used in \cite{Lightman, chluba2} when evolution of the photon occupation number is considered. It is should be noted that behaviour of the left side of Eq. (\ref{tudasuda}) in infrared region depends on the ratio of photon occupation numbers in the right side this equation. Since all $ N_{\textbf{k}\textbf{e}}$ in Eq. (\ref{tudasuda}) are defined by the properties of real source one can conclude that the divergence problem is actually solved by the constraints on the source.

Performing analytical evaluation of cross-section Eq. (\ref{mandl16}) we find that the cross-section of IDC has no infrared divergence in variable $ \omega $ and as a result, the total cross-section becomes finite in contrast to the direct process. However, there is another divergence of the type $ (\omega_1^a\omega_2^b)^{-1} $ ($ a $ and $ b $ are some integers $ \geqslant 1 $) which depends now only on input parameters. Cross-section becomes infinite for the unphysical situation when two incident photons have zero energy. Since the real source of photons always has finite width the frequencies  $ \omega_1$ and $\omega_2 $ in Eq. (\ref{mandl16}) are also finite. Similarly, for a direct process, the solution to the problem of infrared divergence can be resolved taking into account the fact that in under the real physical conditions the photon wave-function can not described by a plane wave.  As a result the physical frequency profile of the scattered radiation for direct as well as inverse DC processes is a function of both the intensity and line shape of the incident photon field \cite{dawson1970}. The monochromatic limit, while it may exist as a mathematical exercise, is not germane to the physical scattering problem. The shape and extent of the external field are fundamental aspects of the problem. 

Thus, we see that the problem of divergence can be solved not only by imposing restrictions on the detector and its resolution, but also by taking into account the spectral characteristics of the source itself, in particular, taking into account the finite interaction time of radiation with an electron. The latter circumstance for the theory of free particles can be taken into account phenomenologically by analogy with the quantum-mechanical description of photon scattering processes on atoms, where singular denominators are regularized by the introduction of atomic level widths \cite{Low, ZSLP-report, Andr}. Then, following \cite{Low} the infrared behaviour $ \omega_1^{-1} $ of the DDC scattering can be naively regularized as follows
\begin{eqnarray}
\label{reg}
d\sigma_{\mathrm{ddc}}\sim \frac{\Gamma}{\omega_1^2+\Gamma^2/4}
,
\end{eqnarray} 
where $ \Gamma=\tau^{-1} $ and $ \tau $ is the interaction time of incident photon with electron. For the inverse process there is no divergence and such regularization is not needed. However the introduction of $ \Gamma $ in this case is also possible and should lead to the correct infrared asymptotic behaviour when two-incident photons are soft.

In conclusion, it should be added that with sufficient intensity of an external astrophysical radiation source, the inverse process can also be of interest in the study and analysis of scattering spectra by electrons \cite{chluba2}. As shown in this work, the cross-section of the inverse process has no infrared features and is determined entirely by the spectral density of incident photons. We leave the application to the study for different astronomical situations for further study. Present calculations represent only the first step towards solving the complete problem of the radiation transfer equation, where incident photons interact with an electron for a finite period of time.

\bibliography{mybibfile} 
\end{document}